\documentclass[journal=nalefd,manuscript=letter]{achemso}
\usepackage{graphicx}
\usepackage{bm}
\usepackage{color}
\usepackage{nicefrac}
\usepackage{amsmath}
\usepackage{verbatim}
\usepackage{longtable}
\usepackage[normalem]{ulem}
\usepackage{appendix}

\newcommand{\ket}[1]{|#1\rangle}
\newcommand{\unit}[1]{\ensuremath{\, \mathrm{#1}}}

\author{Philipp R. Struck}
\email{philipp.struck@uni-konstanz.de}
\author{Heng Wang}
\author{Guido Burkard}
\affiliation[University of Konstanz]{Department of Physics, University of Konstanz, D-78457 Konstanz, Germany}

\title{Nanomechanical read-out of a single spin}


\begin{document}

\begin{abstract}
The spin of a single electron in 
a suspended carbon nanotube can be read out by using its coupling to the 
nano-mechanical motion of the nanotube.  To show this, we consider a single 
electron confined within a quantum dot formed by the suspended carbon
nanotube. 
The spin-orbit interaction induces a coupling between the spin 
and one of the bending modes of the suspended part of the nanotube.  
We calculate the response of the system to pulsed external driving of
the mechanical motion using a  Jaynes-Cummings model.
To account for resonator damping, we solve a quantum master equation, with parameters 
comparable to those used in recent experiments, and show how information of the 
spin state of the system can be acquired by measuring its mechanical motion.  
The latter can be detected by observing the current through a 
nearby charge detector. 
\end{abstract}




In the past few years, advanced manufacturing techniques have spawned new interest in nano-mechanical devices. These are of interest for fundamental research as well for possible applications.
In various experimental setups the cooling of a nanomechanical resonator to its ground state has been achieved.\cite{oconnell2010,chan2011,teufel2011} It is now possible to study quantum features such as the zero-point motion of objects much larger than the atomic scale.\cite{safavi-naeini2012} 
As mechanical motion can be coupled via a wide range of forces, nanomechanical systems have been proposed for various applications such as mass\cite{mamin2001,peng2006}, force\cite{jensen2008}, and motion\cite{teufel2009} sensing. 
This versatility also makes hybrid systems of mechanical devices coupled to other systems possible candidates for various applications in quantum information and communications. 
For example, nanomechanical resonators have been proposed as qubits\cite{fiore2011}, optical delay lines\cite{safavi-naeini2011}, quantum data buses\cite{rabl2010}, and quantum routers\cite{habraken2012} among others.

One of the challenges in building sensitive nano-electromechanical
systems is to control thermal fluctuations. 
Only if the device operates at energies significantly higher than the
thermal energy, one can expect to observe quantum mechanical behavior in equilibrium. 
In this context, carbon is a very promising building material because the
resulting structures are very light and stiff which leads to high resonance frequencies.
Resonators made of suspended carbon nanotubes (CNTs) are of special interest. They can be produced almost free of defects and with radii on the scale of $\approx 1\unit{nm}$ and lengths up to $1\unit{\mu m}$ and in addition their bending mode is easily excited. 
Experimentally it has been demonstrated that frequencies of more than $4 \unit{GHz}$ and quality factors of more than $100,000$ are achievable.\cite{huettel2009, steele2009}

Coupling a localized electron spin to a nanomechanical resonator is of particular interest because such a system may be used as a quantum memory due to the relatively long spin lifetimes.\cite{loss1998}
Carbon based materials with their weak spin-orbit coupling due to the relatively low atomic mass of carbon and the presence of only few nuclear spins is a promising material.\cite{trauzettel2007}
Recently the readout of a single spin of the NV center in diamond has
been used to read out the oscillatory motion of 
the magnetized tip of an atomic force microscope.\cite{kolkowitz2012,rabl2009,balasubramanian2008,bennett2012} 
However, the well-defined local field gradient required for this setup is quite challenging.
CNTs provide an entirely different arena for the coupling of single spins to nanomechanical resonators and 
they can potentially be integrated into scalable (2D) semiconductor structures. 

\begin{figure}[hbpt]
	\begin{center}
		{\includegraphics[width=0.45\textwidth]{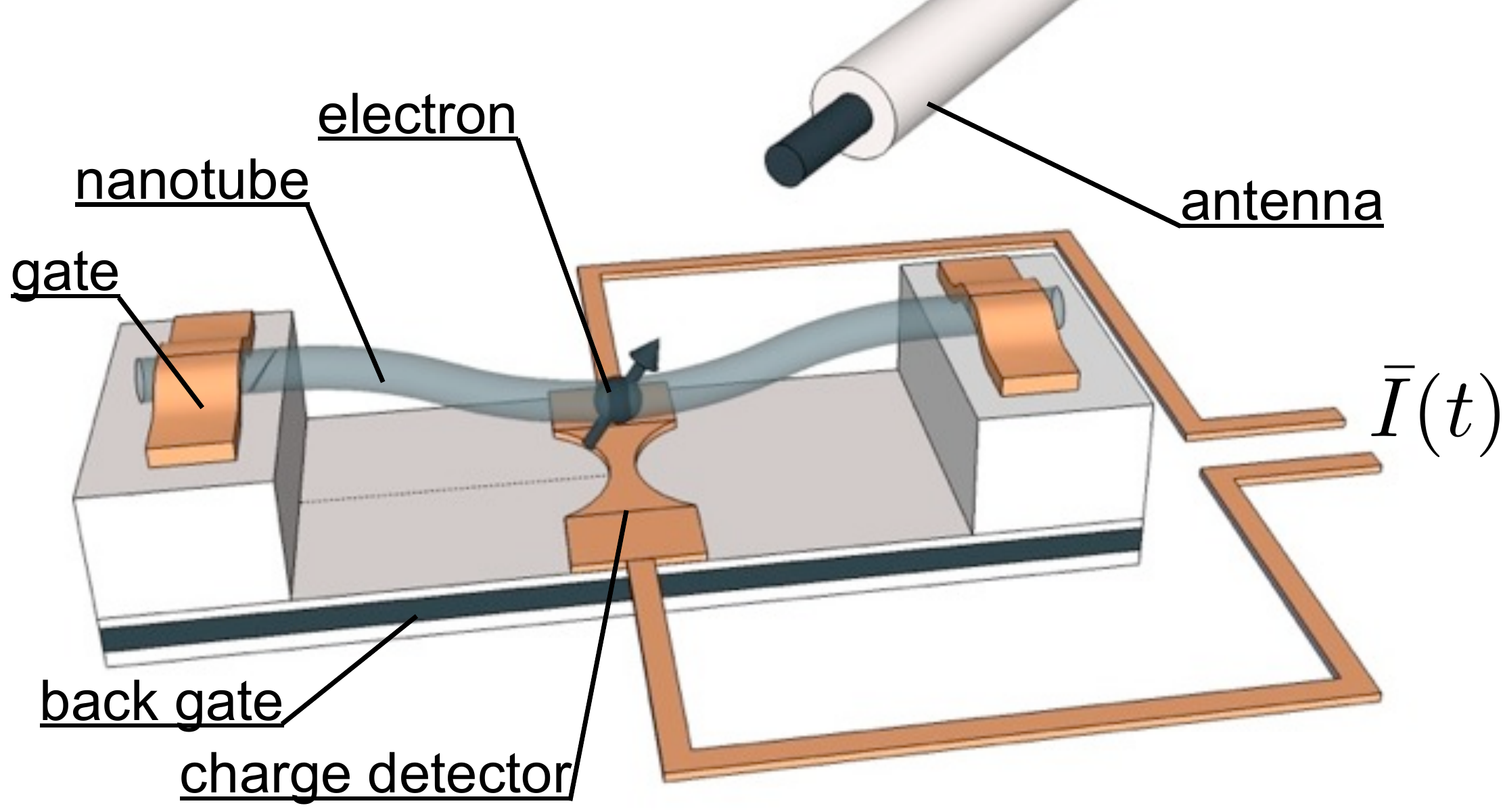}}
	\end{center}
	\caption{Schematic of the proposed setup 
of the carbon nanotube (CNT) resonator with a quantum dot containing a single electron.
	The gates on both sides of the CNT can be electrostatically adjusted to control the number of confined electrons.
	The backgate is used to tune the resonance frequency of the CNT into resonance with the splitting of the Zeeman sublevels of the electron which make up the qubit.
	The charge sensing device in the vicinity to the vibrating nanotube serves as a detection device for changes in the amplitude due to the coupling of the spin to the vibrational motion. 
It can be a quantum point contact (QPC) or a quantum dot (QD) which needs to be operated in a regime where the current depends non-linearly on the gate voltage.
		A current meter is used to measure a nonlinear current $I$ which flows through the device.}
	\label{fig:setup}
\end{figure} 

Here, we propose a method for an all-electrical readout of the state of a single electronic spin. 
We assume that the spin is confined to an oscillator consisting of a doubly clamped,
suspended CNT. 
While previous experiments\cite{huettel2009,steele2009} make use of a charge transport measurement throught the CNT, here we focus on the measurement of the amplitude of the vibrating CNT via a close-by charge sensor, which could be a quantum point contact (QPC) or a quantum dot (QD). 
Those devices have been shown to be sensitive to charge fluctuations
of far less than one elementary charge.\cite{elzerman2003,vandersypen2004}
The coupling of the spin to the vibrational motion allows the read-out of the spin state via measurements of the amplitude of the mechanical vibration.\cite{palyi2012}
This scheme does not require optical access to the probe or time-dependent magnetic fields 
which makes it potentially scalable.



The setup under study is shown in Figure (\ref{fig:setup}). It consists of a CNT suspended over a trench with contacts on
both sides and electrostatic gates responsible for the electron
confinement.
Since the nanotube carries a net charge, a constant (dc) voltage
applied to a back gate can be used to tune the mechanical resonance frequency.
Furthermore, an ac voltage applied to an external antenna or the back
gate can be used to drive the motion of the charged resonator \cite{steele2009,palyi2012}.

In order to obtain a well defined qubit, the fourfold degeneracy due to
the valley and spin degree of freedom of the CNT 
has to be lifted (note that for the localized electron, the sublattice degree of freedom is frozen out).
A magnetic field $B$ parallel to the CNT axis in combination with the
spin-orbit coupling serves this purpose. 
Around a field of $B^* \approx \Delta_{SO}/2\mu_B$ two orthogonal spin
states within the same valley are split by the 
Zeeman energy $\hbar\omega_q = \mu_B (B-B^*)$. Here $\Delta_{SO}$
refers to the intrinsic spin-orbit coupling strength; 
the intervalley coupling $\Delta_{KK'}$ is assumed to be much smaller.
In the following we assume $\Delta_{SO}=370\unit{\mu eV}$ and
$\Delta_{KK'}=65\unit{\mu eV}$ as realistic parameters.\cite{kuemmeth2008}
The corresponding states in the opposite valleys are energetically well seperated.
Thus we can treat the system as an effective two level system.

For the detection of the nano-mechanical oscillation we propose to 
use that the position $x(t)$ of the oscillating 
charged CNT modulates the current through the charge sensor in a (generally) nonlinear way, i.e.
$I(t) = I_0 + I_1 x(t)+I_2 x^2(t)+O(x^3)$.  
This is the case, for example, when a QD is operated at the  
maximum of a Coulomb blockade peak. 
This situation will be assumed for following discussion. 
Furthermore, we assume that the frequency $\omega_p$ of the oscillator is much larger than the tunneling rate through the charge sensing device.
In this case, in lowest order the latter only probes a time averaged squared displacement 
$\overline{X^2}=\frac {1}{\tau} \int_{0}^{\tau}{dt X^2(t)}$
where $X^2(t)=\langle x^2(t) \rangle={\rm
  Tr}({x}^{2}\rho(t))$ with the oscillator density matrix $\rho$.
In the expansion of $I(t)$, only even-order terms in $x$ appear
because $\overline{x^n}=0$ for $n$ odd.
Higher order terms $n>2$ are neglected because they contribute only weakly 
to the current for small displacements $x(t)$. 
Our goal in this paper
is to calculate the time-averaged current $\overline{I}$ through the charge sensing device as a function of
the spin state of the electron.


In the following, we restrict our considerations to one polarization
of the bending mode of the CNT. 
The generalization to two modes is straightforward.
When quantized, the resonator displacement $x$ can be written as $x=\frac{l_{0}}{\sqrt{2}}(a+a^{\dag})$, where
 $a$ and $a^{\dag}$ are phonon creation and annihilation operators and
 $l_{0}$ is the zero-point motion amplitude of oscillator.
In principle there are two ways in which the flexural phonons can couple to the spin. 
At large phonon energies the usual deformation potential is dominant
as it is proportional to $q^2$ where $q$ is the phonon wave number. 
For lower energies however the spin-orbit mediated deflection coupling 
$\propto q$ to flexural phonons dominates.\cite{rudner2010}. 
The resulting coupling strength $g$ will be proportional to both the spin-orbit
coupling $\Delta_{SO}$ and the zero-point amplitude $l_0$.\cite{palyi_suppl_2012}

The system is described by the Jaynes-Cummings Hamiltonian 
\begin{equation}
H= H_0 + H_d
\label{eq:JC_hamiltonian_1}
\end{equation}
with 
\begin{equation}
H_0 = \frac{\hbar\tilde{\omega}
_{q}}{2}\sigma_{z}+\hbar g(a\sigma_{+}+a^{\dag}\sigma_{-})+\hbar \tilde{\omega}_{p}a^{\dag}a, \quad 
H_d = \hbar \lambda(a+a^{\dag}),
\label{eq:JC_hamiltonian_2}
\end{equation}
where $\sigma_{z}$ is a Pauli matrix acting on the spin qubit while
$a$ and $a^{\dagger}$ are the creation and annihilation operators for the phonons. 
The qubit and oscillator frequencies in the rotating frame, $\tilde{\omega}_q=\omega_q-\omega$ and
$\tilde{\omega}_p=\omega_p-\omega$, 
are given as detunings from the driving frequency $\omega$.
The driving strength $\lambda$ is assumed to be  weak, i.e., $\lambda \ll \omega_p$. 
In the Jaynes-Cummings model a rotating-wave approximation is incorporated which 
is valid as long as the driving frequency $\omega$ is comparable to $\omega_q$ and $\omega_p$.


In addition to the unitary evolution we include the damping of
the CNT with a rate  $\Gamma$ which can occur on the same timescale as the read-out via the charge detection device.
The spontaneous qubit relaxation $\gamma=1/T_1$, where $T_1$ denotes the spin relaxation time,  is neglected because of the
very low density of other phonon modes in the vicinity of the bending mode at frequency $\omega_p$.
Previously\cite{palyi2012} we have shown that with state-of-the-art
experimenental techniques the strong-coupling regime, i.e. 
$g\gg \Gamma,\gamma$, is within reach.
The non-unitary dynamics are described by the quantum master equation for the time evolution of the density matrix $\rho$
\begin{equation}
	\begin{split}
	\dot{\rho}=& -\frac{i}{\hbar}[H,\rho] \\
	& +(n_{B}+1)\Gamma(a\rho a^{\dagger} -\frac{1}{2}\{a^{\dagger} a,\rho\}) \\
	& +n_{B}\Gamma(a^{\dagger}\rho a-\frac{1}{2}\{a a^{\dagger},\rho\}).
\end{split}
	\label{eq:master_eqn_finiteT}
\end{equation}
Here $n_{B}=1/(e^{\hbar \omega_{p}/k_{B}T}-1)$ refers to the Bose-Einstein occupation factor of the phonon bath at temperature $T$.
The Lindblad terms $\propto \Gamma$ correspond to emission (absorption) of a phonon to (from) the phonon bath.

\begin{figure*}[hbpt]
		{\includegraphics[width=1.0\textwidth]{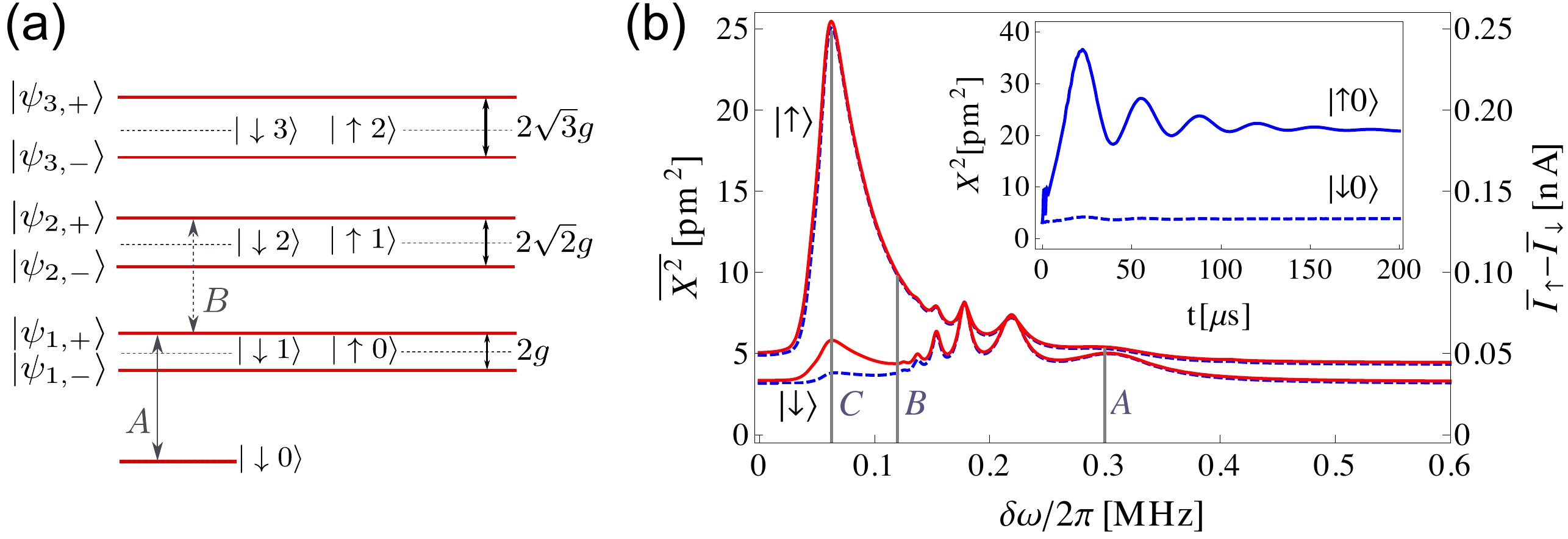}}
	\caption{(a) The eigenenergies of $H_0$ are shown in red. 
	The black dotted lines correspond to the uncoupled,
        i.e. $g=0$, case. Note that the ground state is not modified
        by the coupling $g$. 
	The coupling introduces a splitting $2 g\sqrt{n}$ which is nonlinear in the phonon number $n$.
	(b) The time averaged squared resonator displacement as a function of the detuning of the driving frequency  $\delta \omega=\omega-\omega_p$ for a fixed integration time of $\tau=100\unit{\mu s}$.
	The upper (lower) blue dashed line is the evolution at $T=0\unit{mK}$ of the initial state 
	$|\uparrow0\rangle$ ($|\downarrow0\rangle$). 
	The strongest difference between the amplitudes of the two
        aforementioned initial states is found for a detuning of
        $\delta\omega/2\pi=\pm0.07\unit{MHz}$ (see mark \textit{C}).
	The upper and lower red solid lines correspond to the same
        initial spin states at a temperature of $T=30\ {\rm mK}$. 
	The other parameters are $\lambda/2\pi = 0.04\ {\rm MHz}$, $\omega_p/2\pi =\omega_q/2\pi = 1.5\ {\rm GHz}$, 
	$\Gamma =5\cdot10^{4} {s^{-1}}$, and $g/2\pi =0.3 \ {\rm MHz}$. 
	At $30\unit{mK}$ the thermal energy $k_B T$ is 2.4 times smaller than $\hbar \omega_q$.
	The letter \textit{A} in both figures marks the transition from the ground state to the first excited state with positive detuning. 
	The line marked \textit{B} corresponds to the next transition between states with positive parity. A transition between the states with negative parity requires a negative detuning. 
	In between \textit{A} and \textit{B} only two-phonon transitions can occur.
	The main peak around \textit{C} at $\delta \omega\approx
        0.07\unit{MHz}$ is due to several transitions with the main
        contribution from transtion between eigenstates
        $\ket{\psi_{4,+}} \rightarrow\ket{\psi_{5,+}}$. This is
        consistent with the magnitude of the main peak around $n \sim
        \overline{X^2}/l_0^2 \sim 4$.
	(Inset) The squared amplitude of the CNT is shown as a function of time $t$ for a detuning of the driving frequency of $\delta\omega/2\pi=0.07\ {\rm MHz}$. 
	Within the first $10\,\unit{ns}$ the fast Rabi oscillations with the frequency of the coupling strength $g$ are damped and following oscillations are caused by the driving strength $\lambda$.
	The resonator reaches a steady state after an interaction time $1/\Gamma$.  
	\label{fig:jc-ladder_and_zero_temperature}}
\end{figure*}

While solutions of the quantum master equation
(\plainref{eq:master_eqn_finiteT}) cannot be given in closed form, it
is worthwhile studying the eigenstates of $H_0$ first,
\begin{align}
	|\psi_{n,+}\rangle=\cos\frac{\alpha}{2}|\uparrow n-1\rangle+\sin \frac{\alpha}{2}|\downarrow n\rangle,\\\nonumber
	|\psi_{n,-}\rangle=-\sin\frac{\alpha}{2}|\uparrow n-1\rangle+\cos\frac{\alpha}{2}|\downarrow n\rangle,
\end{align}
for $n\geq 1$ and the special case of the ground state $n=0$:
$\ket{\psi_0}=\ket{0\downarrow}$.
Here, we use the notation $|n \sigma \rangle$ for eigenstates of $H_0$
with $g=0$, i.e., $a^\dagger a|n\sigma\rangle = n|n\sigma\rangle$ and $\sigma_z|n\sigma\rangle = \sigma|n\sigma\rangle$ and $\sigma=\uparrow,\downarrow \equiv\pm1$.
The mixing angle $\alpha$ is defined by $\tan\alpha=\frac{2g\sqrt{n}}{\tilde{\omega}_{p}-\tilde{\omega}_{q}}$. 
In the resonant case $\alpha\rightarrow \pi/2$.
The eigenenergies are 
\begin{equation}
	E_{n,\pm}=\hbar\tilde{\omega_{p}}(n-\frac{1}{2})\pm\hbar\sqrt{\left(\frac{\tilde{\omega_{p}}-\tilde{\omega_{q}}}{2}\right)^2+g^2n}
\end{equation}
for $n\geq1$ and $E_0=-\hbar\frac{\tilde{\omega}_q}{2}$. 
For $n=0$, we have $|\downarrow0\rangle$ as the ground state. 
The energy splitting between adjacent eigenstates (see
Figure (\ref{fig:jc-ladder_and_zero_temperature}a)) in the resonant case $\tilde\omega_{p}=\tilde{\omega_{q}}$ is 
$E_{n+1,\pm}-E_{n,\pm}=\pm\left(\hbar \tilde\omega_p +\hbar g(\sqrt{n+1}-\sqrt{n})\right)$
and 
$E_{n,+}-E_{n,-}=2\hbar g \sqrt{n}$.

Since we are interested in the read-out of the electron spin, we assume an initial state in which the latter is in one of the two $\sigma_z$ eigenstates.
In the case of $T=0$ and an empty QD the osciallator is in its ground state $\ket{0}$ 
with the zero-point amplitude $l_0$.
Directly after loading with an electron the state of the system is $\ket{0\uparrow}$ or $\ket{0\downarrow}$.
At temperatures $T>0$, the distribution of phonons obeys Bose-Einstein statistics and we obtain
\begin{math}
	|\Psi(\sigma) \rangle_{T}= |\sigma\rangle \otimes\frac{1}{Z} \sum_{n=0}^{\infty}e^{-n\hbar\omega_p / k_B T}|n \rangle
	\label{finiteinitial}
\end{math}
as a initial state, where
$\sigma=\uparrow,\downarrow$ and
$Z=\sum_{n=0}^{\infty}e^{-n\hbar\omega_p / k_B T}$ is the partition function.
Experimentally, the state preparation could be performed using
techniques that have been established in conventional semiconductor QDs.\cite{hanson2003} 
Raising one of the gates well above the two Zeeman-split qubit levels allows an electron to hop onto the CNT. 
Provided $g_{\rm el}\mu_B B\gg k_B T,g$, where $g_{\rm el}$ is the
electron g-factor, $\mu_B$ the Bohr magneton, and $B$ the magnetic field, the probability for finding either
Zeeman sublevel to be populated is 1/2. 
A subsequent measurement will then reveal which state was prepared. 
If the measurement is delayed by a time $\tau$, then the population of
the higher-energy spin state will decay as $\sim \exp(-\tau/T_1)$.\cite{elzerman2003}.
To demonstrate the coupling between spin and phonon and to study the read-out this random filling method is sufficient.
For possible application in quantum computing, a controlled state preparation is necessary. 
In this case ferromagnetic leads could be used as has already been
demonstrated for nanotubes deposited on a substrate.\cite{sahoo2005}



We first solve the master equation (\plainref{eq:master_eqn_finiteT})
for the case of $T=0$ 
numerically and calculate the squared oscillator displacement which is
proportional to the current through the charge detection device, in
this case a QD.
We assume that the oscillator frequency of the fundamental bending mode
of the nanotube $\omega_p/2\pi = 1.5\ {\rm  GHz}$ is matched by the
Zeeman splitting $\omega_q/2\pi $ of the electron spin
(the zero point motion for this case amounts to $l_0 = 2.5\unit{pm}$).
At this frequency the ground state energy of the oscillator is 2.4 times smaller than the thermal energy at $T=30\unit{mK}$.
Experimentally, frequencies of suspended nanotubes between $120\unit{MHz}$ and $4.2\unit{GHz}$ have been reported.\cite{huettel2009,chaste2012}
The coupling strength is chosen to be $g/2\pi =0.3 \ {\rm MHz}$ which
is much larger than the damping of the CNT, $\Gamma =0.05  {\rm MHz}$.
Together with the spontaneous relaxation of the qubit $\gamma$ which is negligible the device can be operated in the strong-coupling regime, i.e. $g\gg\Gamma,\gamma$.
The driving strength $\lambda/2\pi = 0.04\ {\rm MHz}$ is chosen to be large enough to compensate for the damping but weak enough not to dominate over the effect of the spin-phonon coupling. 

In the inset of Figure (\ref{fig:jc-ladder_and_zero_temperature}b) the response of the oscillator at temperature $T=0$ is plotted as a function of the driving time $t$ beginning from the initial preparation of the spin state as described above. 
The driving frequency is detuned from the resonant oscillator and qubit frequencies $\omega_p=\omega_q$ by $\delta\omega=\omega-\omega_p$. The value of $\delta\omega/2\pi=0.07\unit{MHz}$ is found to give the largest difference in amplitude with respect to both initial spin orientations as we discuss below.

A considerable difference between different initial spin states in the
squared amplitude (and therefore in the measured current) can be observed. 
For the initial state $\ket{0\uparrow}$, see inset of Figure (\ref{fig:jc-ladder_and_zero_temperature}b), after the inital fast Rabi oscillations with a frequency $g$ disappear (on a timescale $\approx 10 \unit{ns}$), the dynamics are governed by the slower Rabi oscillations at the frequency of the drving strength $\lambda$.
The evolution of the initial state $\ket{0\downarrow}$ shows only little effect of the driving. 
This can be explained by the nonlinearity introduced by the (strong)
spin-phonon coupling in 
the Jaynes-Cummings Hamiltonian (\plainref{eq:JC_hamiltonian_1}). 
When driving with frequencies detuned less than
$(\sqrt{2}-\sqrt{1})g\approx0.12\unit{MHz}$ from resonance, 
the probability to leave the ground state is much less than for all other states.

Typically, the charge detector is too slow to follow the instantaneous
motion of the resonator, and it will thus detect an averaged signal
$\overline{I} \propto \overline{X^2}$ caused by the vibrating charge.
In the main part of Figure (\ref{fig:jc-ladder_and_zero_temperature}b) we plot the integrated averaged squared resonator displacement $\overline{X^2}$ as a function of the detuning of the driving frequency $\delta\omega$. 
As an integration time we choose $\tau=100\unit{\mu s}$;
however, any integration time $\tau$ of more than $30\unit{\mu s}$ was
found to yield a useful signal.

The peaks can be explained with the help of the spectrum (Figure (\ref{fig:jc-ladder_and_zero_temperature}a)) of the Jaynes-Cummings Hamiltonian (\plainref{eq:JC_hamiltonian_2}). Whenever the driving frequency hits a resonance, energy is absorbed by the oscillator which results in an increased amplitude.
While only positive detuning are shown here, note that the result for negative detunings is exactly the same.
The peak at $\delta\omega/2\pi= 0.3\ {\rm MHz} = g/2\pi$, denoted by
\textit{A}, for initial state $|\downarrow 0\rangle$ for example is
caused by a strong resonant coupling between the ground state
$|\downarrow 0\rangle$ and the dressed state  $|\psi_{1,+}\rangle$.
Note that the smaller peaks between \textit{A} and \textit{B} are due to two-phonon processes
which are less pronounced because of the weak driving. 
The main peak \textit{C} centered around $0.07\unit{MHz}$ is due to transitions between states of equal parity, i.e. $|\psi_{n,+}\rangle \rightarrow |\psi_{n+1,+}\rangle$ with $n\geq 1$.
The line \textit{B} marks the transitions  $|\psi_{1,+}\rangle \rightarrow |\psi_{2,+}\rangle$ which corresponds to a detuning of the driving frequency of $\delta\omega\approx 0.12\unit{MHz}$. 
The main contribution to the large peak \textit{C} comes from the
transition $|\psi_{4,+}\rangle \rightarrow |\psi_{5,+}\rangle$ 
which corresponds to $0.07\unit{MHz}$.
This is consistent with the peak hight of $\overline{X^2} \approx 25\,\unit{pm}$ which in turn corresponds to a phonon number 
$n\sim \overline{X^2}/ l_0^2\sim 4$.
The individual transitions cannot be resolved because the lines are broadened due to the finite damping of the CNT. Note however that a finite damping is essential to allow for non-resonant transitions which enable population of higher state. 
The large energy gap between the ground state and the first excited state due to the nonlinearity of the spectrum is the reason why the two spin states $\ket{\uparrow}$ and $\ket{\downarrow}$ can be resolved.

In Figure (\ref{fig:jc-ladder_and_zero_temperature}b) the solid red lines show
the integrated amplitude in the case of a
 finite temperature of $T=30\unit{mK}$. 
Qualitatively, the same features are observed as for $T=0$ but in this case
the integrated amplitude for the initial 
state $\ket{\psi(\downarrow  )}$ is larger because of the thermal
distribution of oscillator in $\ket{\psi(\downarrow)}$,
containing a small admixture of $\ket{1\downarrow}$ and higher states, in addition to $\ket{0\downarrow}$.
We also note that the width of the peak is only little affected by the
finite temperature and is still goverend by the damping of the CNT.


To achieve the most efficient read-out scheme we strive to increase
the contrast in the integrated amplitude between spin up and down. 
A way to achieve this is to change the driving frequency as a function
of time. 
This compensates for the fact that the frequencies of higher transitions are smaller than those of the lower transitions. 
From Figure (\ref{fig:jc-ladder_and_zero_temperature}a) we see that the
transition frequencies between states of 
same parity scales with $g\left(\sqrt{n+1}-\sqrt{n}\right)$ (see above).
For the sake of simplicity in the following we change the
frequency once to demonstrate the principle, but this method could be
optimized to further increase the contrast.
To achieve the maximum contrast, i.e. the difference between the integrated amplitudes for the two initial states with different spin orientations, we switch the frequency at the point right before the squared amplitude reaches its maximum.
The parameters used are the same as in the case of continuous driving. 
The time at which the driving frequency is switched from $1500.070 \unit{MHz}$ to $1500.036 \unit{MHz}$  is $19 \unit{\mu s}$. 

\begin{figure}[hbpt]
	\begin{center}
		{\includegraphics[width=0.5\textwidth]{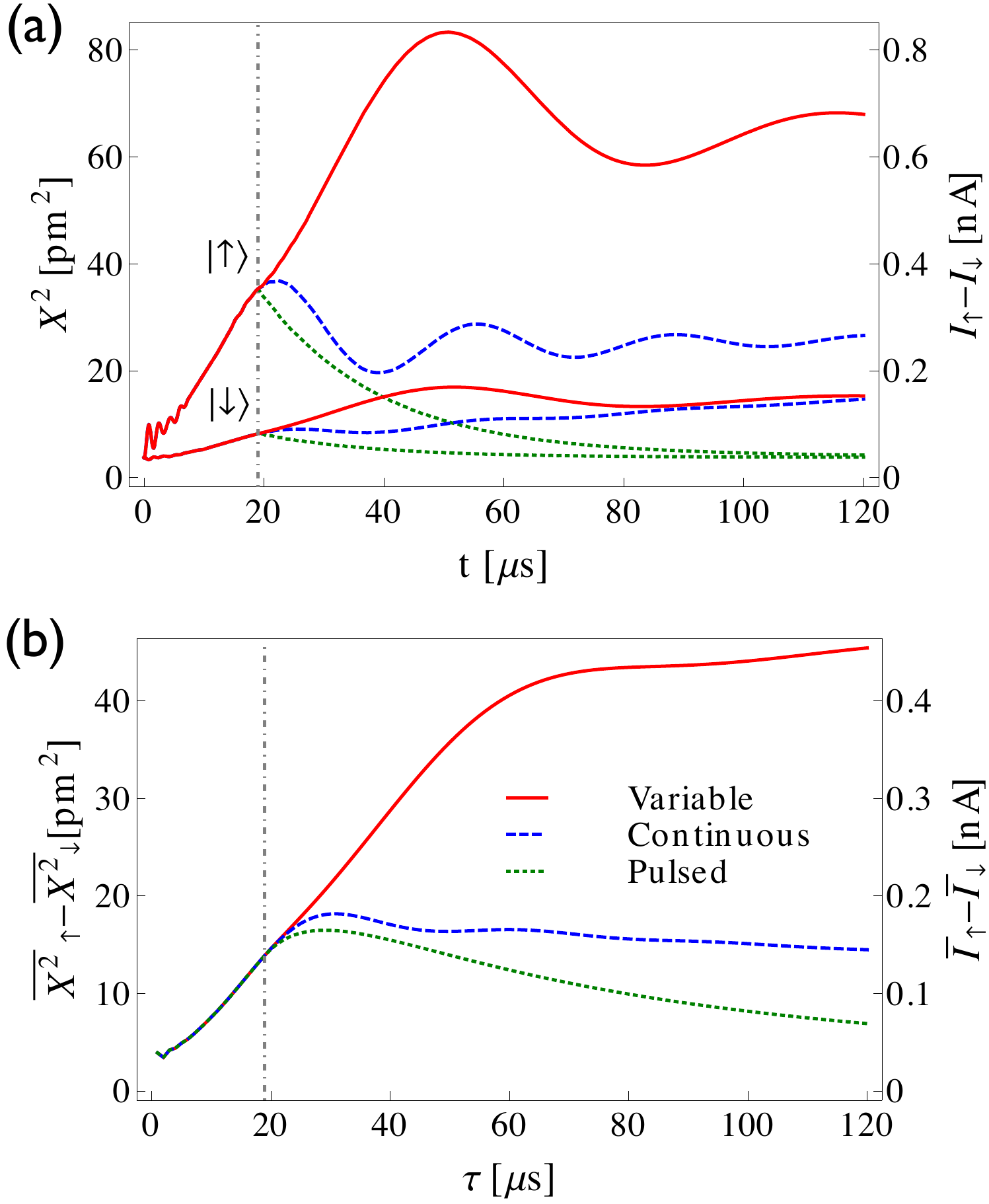}}
	\end{center}
	\caption{(a) The squared amplitude as a function of the elapsed time $t$ is shown for the cases of continuous (dashed blue), pulsed (dotted yellow), and variable (solid red) driving. 
	The same parameters are used as in Figure (\ref{fig:jc-ladder_and_zero_temperature}).
	The driving frequency $\omega$ is changed at the time $t=19 \unit{\mu s}$ right before the amplitude is largest.
	(b) The three curves corredspond to the same modes of driving. Each curve shows the difference between the two corresponding squared amplitudes (on the left $y$-axis) and current (on the right $y$-axis) integrated over an interval given by the time $\tau$ in the $x$-axis, which is proportional to the average current $\overline{I}_{\uparrow}-\overline{I}_{\downarrow}$ through the QD.
	The variable driving scheme results in an increase in current by a factor of about 2.}
	\label{fig:variable_driving}
\end{figure} 

The result is shown in Figure (\ref{fig:variable_driving}a).
We clearly observe an increase in the squared amplitude of the resonator for the case of the initial state $\ket{\psi \uparrow}$ while the evolution of the other initial state is barely affected.
To demonstrate the effectiveness of the driving scheme, we also plot
the response of the system when the driving is switched off 
at $19 \unit{\mu s}$. 
As expected, this causes the current to decay exponentially with a rate $\Gamma$.
This pulsed driving results in lower contrast as long as the integration time $\tau$ is on the same order as the time scale $1/\Gamma$ given by the damping and not considerably longer.

To demonstrate the feasiblility of our proposal we use a simple model of a QD capacitively coupled to the vibrating charge distribution on the CNT to derive an estimate for the current fluctuations.
In the linear (low-bias voltage) regime, the current is given by $I(V_{\mathrm{gate}})=V_{sd}G(V_{\mathrm{gate}})$.
The conductance $G(V_{\mathrm{gate}})$ is assumed to be a Lorentzian
as a function of the gate voltage which, 
due to capacitive coupling of QD and charged CNT, in turn depends on
the squared average displacement $\overline{X^2}$ of the CNT. 
The average displacement $\overline{X}$ itself is zero.
To get an estimate of the capacitance we use a simple toy model in
which both the QD 
and he charged CNT are tubes of finite length $l$ and equal diameter.
While this is a relatively crude model, it gives a lower limit as we show in the Appendix.
Expanding both the capacitance as well as the current $\overline{I}$ up to second order in $\overline{X}$ and  
in the limit of a tube radius much smaller ($1\unit{nm}$) than the distance between the CNT and the QPC ($d=50\unit{nm}$) we obtain 
\begin{equation}
	\overline{I(X)} = I_0+I_2 \overline{X^2} = V_{sd}G_0 \left(
                              1- \frac{4 e^2}{\pi^2 \Delta V^2 \varepsilon_0^2 l^2 d^2} \overline{X^2}\right),
	\label{eq:current}
\end{equation}
where $G_0$ is the maximum conductance $G_0=2e^2/h$ and $\varepsilon_0$ the vacuum permittivity. 
We assume $\Delta V=0.1\unit{mV}$\cite{beenakker1991}, $V_{sd}=250\unit{\mu V}$,\cite{elzerman2003} and $l=100\unit{nm}$.
With these values we find a constant background current $I_0\approx 20\unit{nA}$. 
The change in current for a maximum average squared amplitude of
$83\unit{pm}$ (cf.~ Figure (\ref{fig:variable_driving}a)) 
is $I_2\overline{X^2}\approx -0.83\unit{nA}$, which corresponds to a fluctuaction in the current of about $4.2\%$ to be detected.
In Figure (\ref{fig:variable_driving}b) we show both the integrated difference in average squared amplitude as well as the corresponding current for the two different initial states for the three kinds of driving as.


In conclusion, we have theoretically shown how the spin state of an
electron coupled to a nanomechanical resonator can be read out through
an adjacent charge sensor. 
Here, we have studied the case where the charge sensor is a quantum
dot, but similar considerations hold for the case of a quantum point
contact, with a slightly altered value of $I_2$.
The read-out scheme is all-electrical and requires no optical access to the device
nor any magnetic field gradients which is a promising in view of its
potential for scaling to many qubits. 
We presented a numerical study of the read-out using realistic experimental parameters. 
We use a simplified model of the QD and the system geometry to
demonstrate that with currently available experimental
techniques  spin read-out via the mechanical motion should be possible.
For small detuning of the driving frequency we observe a maximum contrast between initial states with spin-up and spin-down. 
The contrast decreases with increasing temperature but is still significant at dilution refrigerators temperatures.
We have also shown that more elaborate driving schemes in which the
driving frequency is a function of time can lead to a large increase
in contrast. Additional refinements may be used to further increase
the amplitude of the resonator and thus the sensitiveity and speed of
the spin readout.

We thank Andras Palyi for fruitful discussions.
This work was supported by DFG under the programs FOR 912 and SFB 767.

\newpage
\begin{appendix}
\section{Appendix: Capacitive coupling of CNT and QD}

Based on the physical properties of the CNT resonator, we have calculated its
spin-dependent mean squared resonator amplitude $\overline{X^2}$ after external driving. 
We propose to read out the spin via the influence of the oscillating
resonator on the current through a carge sensing device in the vicinity of the CNT. 
Here, we provide a simplified electrostatic model of the system which we use to estimate the capacitive coupling.
In the following we assume the charge sensing to be performed by a
QD (similar considerations hold for the case of a quantum point contact).

The measured curent is given by $I(V_{\mathrm{gate}})=V_{sd}G(V_{\mathrm{gate}})$ with a fixed bias voltage $V_{sd}$ and a conductance $G(V_{\mathrm{gate}})$. 
The latter depends on the gate voltage $V_{\mathrm{gate}}$ and it is modulated by the oscillating charged CNT due to the capacitive coupling to the QD. 
We assume the QD to be operate in the low-bias regime so that only a single energy level is within the bias window. In this case the (two-terminal) conductance $G$ as a function of the gate voltage is given by a Lorentzian 
\begin{equation}
	I(V_{\mathrm{gate}})=V_{sd}G(V_{\mathrm{gate}})=V_{sd}G_0\frac{\Delta V^2}{\Delta V^2 +4 (V_{\mathrm{gate}}-V_0)^2}
	\label{eq:a_lorentzian}
\end{equation}
where  the conductance at resonance is one unit of the quantum of
conductance, $G_0=2e^2/h$, with a factor of 2 accounting for the spin. 
The width of the Coulomb peak around the center voltage $V_0$ is
assumed to be $\Delta V=0.1\unit{mV}$.\cite{beenakker1991} 
At $30\unit{mK}$ a broadening of the conductance due to temperature is therefore negligible.
The gate voltage which depends on the displacement $x$ in the case of a single charged CNT is simply given by $V_{\mathrm{gate}}=-e/C(d+x)$. 

To find a simple analytical expressions for the the capacitance for the
proposed setup, we use a simple toy model which captures the most 
essential features of our proposed setup.
The shapes of both the charge distribution on the CNT as well as the conducting
island of the QD are approximated as tubes of length $l$ and diameter $a$. 
In the case of $d=50\unit{nm}$, $a=2\unit{nm}$, and $l=100\unit{nm}$, i.e., $a<d<l$,
the capacitance is given by $C(d+x)=\frac{\pi\varepsilon_0 l}{\ln (\frac{d+x}{a})}$ where $\varepsilon_0$ is the vacuum permittivity.
A plate capacitor of equal area of the plates would have a smaller capacitance for all values of $d$ as long as the condition $d\gg a$ holds for the two wires. 
Because in the proposed setup the conducting island of the QD resembles more a plate than a wire our toy model underestimates the actual capacitance and the 
result should be taken as an estimate of a lower limit.

To get a simpler expression we want to expand Eq.~(\plainref{eq:a_lorentzian}) up to second order in $x$. 
We show that this is a good approximation by comparing the fluctuation
in gate voltage caused 
by the oscillating CNT $\delta V_{\mathrm{gate}}$ to the width of the Coulomb peak $\Delta V$.
As shown in Figure (\ref{fig:variable_driving}b), the maxium $x$ is less than
$10\unit{pm}$. 
For this value we find $\delta V_{\mathrm{gate}}=-e\left(\frac{1}{C(d+x)}-\frac{1}{C(d-x)}\right)\approx 0.02\unit{mV}$ 
which is much smaller than $\Delta V=0.1\unit{mV}$.
Expanding Eq.~(\plainref{eq:a_lorentzian}) up to second order in $x$ yields
Eq.~(\plainref{eq:current}),
where $I_0$ is the current with the CNT at rest. 
For a bias voltage of $V_{sd}=250\unit{\mu V}$ we find $I_0\approx 20\unit{nA}$. 
The change in current for a maximum average squared amplitude of $83\unit{pm}$ (cf.~ Figure (\ref{fig:variable_driving}a)) is $I_2\overline{X^2}\approx -0.83\unit{nA}$.
In Figure (\ref{fig:variable_driving}b) we plot the difference between the
currents for the states with initial spin orientation $\uparrow$ and
$\downarrow$ where the background current $I_0$ cancels.
\end{appendix}

\end{document}